# AXION BEAMS AT HERA?

K. PIOTRZKOWSKI [†]

*Département de physique (CP3), Université catholique de Louvain,*
*Chemin du Cyclotron 2, B-1348 Louvain-la-Neuve, Belgium*

If the recently observed anomaly in the PVLAS experiment is due to the axion, then the powerful beams of synchrotron photons, propagating through high magnetic field of the HERA beamline, become strong axion sources. This gives a unique opportunity of detection of the axion-photon interactions by installing a small detector in the HERA tunnel, and to corroborate the axion hypothesis within a few days of running.

## 1. Introduction

Axion hypothesis has remained one of the big open questions in particle physics for almost 30 years. Conceived initially to resolve the problem of strong CP violation in QCD, the axion has gradually become also one of the candidates for cold dark matter in the universe.

Recently, the PVLAS collaboration has observed anomaly in behavior of the linearly polarized laser beams propagating through a strong magnetic field in vacuum – the observed miniscule rotation of the polarization plane (dichroism) can be interpreted as due to the axion production in the Primakoff process (i.e. via photon-photon fusion) [1]. This has been further confirmed by PVLAS preliminary evidence of induced non-zero ellipticity (birefringence) of such a laser beam [2]. Using the PVLAS published and preliminary results, as well as constraints from other experiments, the axion mass and its coupling to photons $g_{\gamma\gamma A}$ have been estimated to be about 1 meV and $2\times10^{-6}\,\text{GeV}^{-1}$, respectively [3].

Another, very complementary technique of detecting photon-axion interactions, is based on the so-called photon regeneration, or *light shining through walls* concept [4]. It requires a strong source of axions and an isolated detector next to a zone of strong magnetic field – as argued below, such a photon regeneration experiment could be performed at HERA with modest effort and at low cost, and would provide an important verification of the PVLAS axion interpretation.

---

[†] Mail: krzysztof.piotrzkowski@fynu.ucl.ac.be.

## 2. Axion production at HERA

In axion searches based on the photon regeneration technique, a very intense beam of photons propagates through the region of strong magnetic field where a small fraction of photons is converted to axions via the coherent photon-photon fusion process. Then, the produced axions, very weakly interacting with ordinary matter, penetrate through a thick shielding to be detected by observation of photons regenerated in the inverted process. The master formula for the regeneration experiments reads:

$$N = 0.6\, N_0\, (g_{\gamma\gamma A}/10^{-6}\ \text{GeV}^{-1})^4\, (BL_1/100\ \text{Tm})^2 (BL_2/100\ \text{Tm})^2, \quad (1)$$

where N is the regenerated photon rate per second, $N_0$ is the initial photon flux in the units of $10^{17}$ photons/s, $BL_1$ and $BL_2$ are the bending powers of the magnets used for the initial photon-axion conversion and final axion-photon regeneration, respectively. It is assumed that direction of magnetic field is parallel to the photon electric field, and that the energy of initial photons $\omega$ is much larger than the axion mass $m_A$, so that $Lm_A^2/2\omega \ll 1$.

HERA electron[*] beam is a very strong source of synchrotron radiation (SR), especially in straight sections of the *ep* experiments, H1 and ZEUS, where very strong magnetic fields are needed to separate the electron and proton beams. The electrons of 27.6 GeV are deflected in the horizontal plane just before and after the interaction point (IP) with the bending radius of about 400 m. This corresponds to the synchrotron radiation critical energy of about 100 keV and angular dispersion, with respect to the instantaneous direction of the electron momentum, of 0.511 MeV/27.6 GeV ≈ 20 μrad. The lateral beam size at these locations is about 1 mm and the beam angular dispersion is about 30 μrad. Axion detection should cover the radiation directions within ±0.5 mrad w.r.t. the collision axis in the horizontal plane, where this strong bending takes place and for which the SR photons will travel through long regions of strong magnetic fields, as shown in Fig. 1, where layout of this area is sketched.

The average number of synchrotron photons radiated within 1 mrad per beam electron is about 0.57, therefore for a typical beam population of $5\times10^{12}$ electrons and its revolution frequency of 47.3 kHz one obtains the SR flux in this forward direction of $1.35\times10^{17}$ photons per second. The SR photons are polarized linearly in the horizontal plane at about 75% on average [5]. Therefore, the principal contribution to the photon-axion conversion will be due to horizontal magnetic dipole field, used at HERA to move vertically the proton beam in the straight sections – its bending power $BL_1$ is about 18 Tm, and is located between 65 m and 79 m from the IP (see Fig.1). Using Eq. 1, the rate of produced axions in one straight section, assuming $g_{\gamma\gamma A} = 2\times10^{-6}$ GeV$^{-1}$, is estimated to be $3.8\times10^{7}$ per second. In such a case, one can indeed talk about axion beams at HERA, with the significant intensity and very specific angular and temporal characteristics, which could be utilized to observe axion-photon interactions. The contribution of vertically polarized photons converting in the vertical magnetic dipole field is less than a couple of percent of the horizontal contribution, and is neglected here. In principle, one could consider also locations outside the straight sections, in the HERA arcs, but there only relatively weak vertical magnetic dipole field is present (for the electron beam), so the synchrotron radiation is weaker and conversions are much less effective.

---

[*] In this paper "electron" refers both to electrons and positrons.

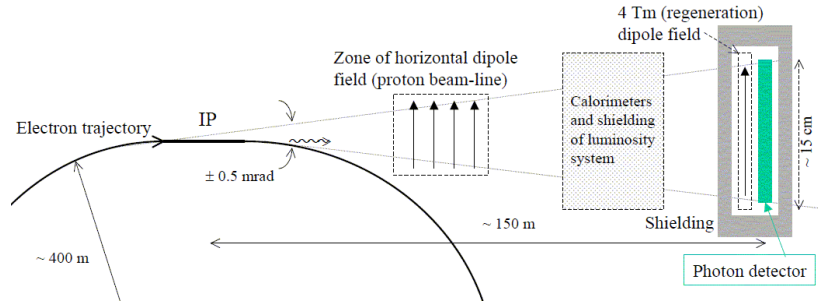

Figure 1: Sketch of the axion production at HERA and of a possible layout of the axion experiment.

Therefore, it is proposed to install the regeneration detector in the direction of electron beam at the interaction point (IP). Assuming a bending power $BL_2$ of the regeneration detector of 4 Tm, one can obtain using the master equation above, the estimated rate of more than two regenerated photons per hour, assuming the PVLAS value of the photon-axion coupling. This should allow for confirmation of the PVLAS axion signal within a couple of days, assuming no large backgrounds, and high regenerated photon detection efficiency.

### 3. Axion detection

It is proposed to install axion detectors in one (or both) of the HERA straight sections, more than 100 m upstream the proton beam, behind the presently running luminosity detectors of H1 and ZEUS. The detectors should register efficiently photons in the 10 keV to 1 MeV energy range, and be surrounded by thick lead shielding, providing protection against large HERA backgrounds. In addition, active high-energy background suppression can be done by a set of scintillator veto counters inside the shielding and around photon detectors. The detector should be located so as to separate maximally in time the signal and the proton beam induced background – for example, at the distance of about 150 m from the IP. The lateral size of the `axion beam' at this position is about 15 cm horizontally and 10 cm vertically defining the minimal detector transversal size.

The crucial characteristic of the axion signal is its arrival time. If measured with respect to the HERA beam-crossing clock of 10.41 MHz it is very narrowly distributed with *rms* of 25 ps, which is dominated by fluctuations of the arrival time of beam electrons at the IP, or equivalently by the electron bunch length of about 8 cm. The photon detectors should be therefore very fast – to be capable of measuring signals for each HERA electron beam crossing every 96 ns and providing an excellent timing resolution, comparable to 25 ps. The axion signal will be then seen as an enhancement of events with timing compatible within resolution with propagation at the speed of light, directly from the IP (i.e. without scattering). A very good quality HERA reference clock signal is essential for that purpose. One should note, for example, that an allowed 100 ps time interval corresponds to a light-path of 30 mm, which means that all the scattered synchrotron photons leaking through the shielding should be strongly suppressed, at the level of 1:1000, as the time interval between beam crossings is 96 ns. The shielding must be therefore thick enough to absorb effectively all the synchrotron photons coming directly from the IP – to large extend this is already achieved by the present luminosity systems, comprising calorimeters and a thick lead shielding, installed some 110 m from the IP, in front of the proposed axion detector location.

Given the required high detection efficiency and very good time resolution, plastic scintillators read out by fast photomultiplier tubes (PMTs) are interesting candidates for the photon detector. Recently, timing resolution better than 30 ps has been reported for the near-beam detectors of similar size at RHIC [6]. However, due to low photon energies the expected number of photoelectrons is very limited in such a detector at HERA, so use of single photon sensitive PMTs utilizing microchannel plates (MCP) should be considered. In this case, the transit time spread (TTS) in a PMT can be very small, below 30 ps, even for events with only one photoelectron. In addition, such photomultipliers are immune to high magnetic field. It is therefore proposed to use large area (2 inch × 2 inch) multi-anode Burle MCP-PMTs with the TTS of about 40 ps [7]. If an efficient shielding could reduce the rate of background photons below a 10 Hz level, the timing cuts should further reduce the backgrounds to the level comparable to the axion signal. Then, indeed already within a few days of HERA running the axion observation might be possible. In addition, the measured photon energies could provide another handle for the signal-background separation. Understanding of the actual separation power of this variable requires however much more detailed studies.

## 4. Summary and outlook

If the PVLAS anomaly is due to the light axion, then HERA becomes a strong axion source with very distinctive angular and temporal distributions. This gives an excellent opportunity of making a low cost photon regeneration experiment in the unique energy range around 100 keV. Further studies are needed to see if a small regeneration dipole could be installed in the HERA tunnel, including the necessary detector shielding. The required photon detection efficiency and time resolution, as well as the expected noise levels, should be demonstrated by a detector prototype. Eventually, reliable calculations of the synchrotron radiation backgrounds inside the shielding are mandatory.

Finally, it should be noted that conditions at HERA are exceptional in the context of axion production. Indeed, HERA provides a unique combination of very high flux of synchrotron radiation and powerful horizontal dipole fields due to the proton beamline. None of existing or planned electron machines, as B-factories or the ILC, offers similar potential.

**References**

1. E. Zavattini *et al*. [PVLAS Collab.], Phys. Rev. Lett. **96** (2006) 110406.
2. M. Gastaldi, on behalf of the PVLAS Collaboration, talk at ICHEP'06, Moscow, http://ichep06.jinr.ru/reports/42_1s2_13p10_gastaldi.ppt.
3. M. Ahlers *et al*., hep-ph/0612098.
4. P. Sikivie, Phys. Rev. Lett. **51** (1983) 1415 [Erratum – ibid. **52** (1984) 695]; A.A. Anselm, Yad. Fiz. **42** (1985) 1480; M. Gasperini, Phys. Rev. Lett. **59** (1987) 396; K. van Bibber *et al.*, Phys. Rev. Lett. **59** (1987) 759.
5. J.D. Jackson, *Classical Electrodynamics*, John Wiley & Sons, New York, 1998.
6. W.J. Llope *et al.*, *Nucl. Inst. Meth. A* **522** (2004) 252.
7. See Burle's web site, http://www.burle.com.